\documentclass[conference]{IEEEtran}
\pdfoutput=1

\IEEEoverridecommandlockouts
\usepackage{amsmath,amssymb,amsfonts}
\usepackage{algorithmic}
\usepackage{graphicx}
\usepackage{textcomp}
\usepackage{xcolor}
\def\BibTeX{{\rm B\kern-.05em{\sc i\kern-.025em b}\kern-.08em
    T\kern-.1667em\lower.7ex\hbox{E}\kern-.125emX}}

\usepackage{subfig}
\usepackage[nolist,nohyperlinks]{acronym}
\usepackage{siunitx}
\usepackage{cite}
\usepackage{hyperref}

\hypersetup{
    colorlinks=false, 				%
    pdfborder={0 0 0}, 		%
    linkbordercolor={1 1 1},  	%
    citebordercolor={1 1 1},   %
    urlbordercolor={1 1 1}     %
}

\makeatletter
\newcommand\footnoteref[1]{\protected@xdef\@thefnmark{\ref{#1}}\@footnotemark}
\makeatother

\usepackage{caption}  
\renewcommand{\baselinestretch}{.97} 
\usepackage{microtype}

\newcommand{\Conv}{%
  \mathop{\scalebox{1.5}{\raisebox{-0.2ex}{$\circledast$}}
  }
}
\DeclareMathOperator*{\argmax}{arg\,max}	%

\acrodef{MC}[MC]{synthetic molecular communication}
\acrodef{EM}[EM]{electromagnetic wave}
\acrodef{CVS}[CVS]{cardiovascular system}
\acrodef{CIR}[CIR]{channel impulse response}
\acrodef{FWHM}[FWHM]{full width at half maximum}
\acrodef{IoBNT}[IoBNT]{Internet of Bio-Nano Things}
\acrodef{SPION}[SPION]{superparamagnetic iron-oxide nanoparticle}
\acrodef{SNR}[SNR]{signal-to-noise ratio}
\acrodef{TX}[TX]{transmitter}
\acrodef{RX}[RX]{receiver}
\acrodef{LBVN}[LBVN]{linear branched vessel network}
\acrodefplural{SPION}[SPIONs]{superparamagnetic iron-oxide nanoparticles}
\acrodefplural{LBVN}[LBVNs]{linear branched vessel networks}
\acrodefplural{SNR}[SNRs]{signal-to-noise ratios}
\acrodefplural{RX}[RXs]{receivers}
\acrodefplural{GFP}[GFPs]{green fluorescent proteins}
\acrodefplural{CIR}[CIRs]{channel impulse responses}

\begin{document}

\title{Molecular Signal Reception in Complex Vessel Networks: The Role of the Network Topology\vspace*{-3mm}
\thanks{This work was funded in part by the Deutsche Forschungsgemeinschaft (DFG, German Research Foundation) -- GRK 2950 -- ProjectID 509922606, and in part by the German Federal Ministry of Education and Research (BMBF) through Project IoBNT.}
}

\author{
\IEEEauthorblockN{Timo Jakumeit$^1$, Lukas Brand$^1$, Jens Kirchner$^2$, Robert Schober$^1$, and Sebastian Lotter$^1$}\\[-0.4cm]
\IEEEauthorblockA{$^1$\small Friedrich-Alexander-Universit\"at Erlangen-N\"urnberg, Erlangen, Germany\\
$^2$Fachhochschule Dortmund, University of Applied Sciences and Arts, Dortmund, Germany}
}

\maketitle

\begin{abstract}
The notion of \ac{MC} refers to the transmission of information via molecules and is largely foreseen for use within the human body, where traditional \ac{EM}-based communication is impractical.
\ac{MC} is anticipated to enable innovative medical applications, such as early-stage tumor detection, targeted drug delivery, and holistic approaches like the \ac{IoBNT}. 
Many of these applications involve parts of the human \ac{CVS}, here referred to as \textit{networks}, posing challenges for \ac{MC} due to their complex, highly branched vessel structures.
To gain a better understanding of how the topology of such branched vessel networks affects the reception of a molecular signal at a target location, e.g., the network outlet, we present a generic analytical end-to-end model that characterizes molecule propagation and reception in \acp{LBVN}. 
We specialize this generic model to any \ac{MC} system employing \acp{SPION} as signaling molecules and a planar coil as \ac{RX}. 
By considering components that have been previously established in testbeds, we effectively isolate the impact of the network topology and validate our theoretical model with testbed data. 
Additionally, we propose two metrics, namely the \textit{molecule delay} and the \textit{multi-path spread}, that relate the \ac{LBVN} topology to the molecule dispersion induced by the network, thereby linking the network structure to the \ac{SNR} at the target location. 
This allows the characterization of the \ac{SNR} at any point in the network \textit{solely based on the network topology}. 
Consequently, our framework can, e.g., be exploited for optimal sensor placement in the \ac{CVS} or identification of suitable testbed topologies for given \ac{SNR} requirements.
\end{abstract}

\acresetall


\vspace*{-2mm}
\section{Introduction}\vspace*{-1mm}
The concept of \ac{MC} stands in contrast to traditional \ac{EM}-based communication and refers to the transmission of information via signaling molecules. 
The primary application domain of \ac{MC} is the human body, where the liquid environment impedes \ac{EM}-based communication between nanoscale devices owing to poor signal propagation conditions.
Here, \ac{MC} is anticipated to facilitate the development of innovative medical procedures by enabling the exchange of information between biological and synthetic systems.

Many in-body applications are envisioned to operate in the \ac{CVS}, which serves as a central distribution network within the body.
Examples include tumor detection, whereby the presence of cancerous tissue is detected by nanosensors that measure elevated concentrations of biomarkers in the vicinity of the tumor~\cite{Mosayebi2019}. 
Similarly, tumor targeting involves functionalized nanoparticles that target the diseased region and deliver drugs for the destruction of malicious cells, thereby minimizing adverse side effects elsewhere in the body~\cite{Simo2024}.
Another example is the \ac{IoBNT}, in which gateways receive molecular information inside the body, e.g., along the \ac{CVS}, and then transmit the information via \ac{EM} signals to an external computational unit, which performs processing and evaluation. 
These gateways work bidirectionally, i.e., they can also receive \ac{EM} signals from the computational unit and convert them into a molecular signal for internal transmission, thereby seamlessly integrating both communication paradigms to enable, e.g., health monitoring and personalized treatment~\cite{Akyildiz2015}.

The implementation of such applications is challenging due to the highly branched nature of the \ac{CVS}, which comprises vessels, \textit{bifurcations} (one vessel splitting up into multiple vessels), and \textit{junctions} (multiple vessels joining into one), resulting in complex dynamics of signaling molecule transport. 
To gain a deeper understanding of molecule propagation in the \ac{CVS}, several theoretical studies have been conducted. In~\cite{Chahibi2013}, a channel model for branched vessel structures based on \acp{CIR} is proposed. 
The authors focus on molecule propagation in arterial tree structures, i.e., on bifurcations, with less consideration given to vessel junctions. 
A method for early cancer detection that uses mobile nanosensors traversing the \ac{CVS} is introduced in \cite{Mosayebi2019}.
The model is developed under the assumption of a quasi-steady state signaling molecule concentration, which in many systems is not applicable due to the time variance of the molecule release.
In~\cite{Pal2023}, a framework for estimating the required dosage of cellular vaccines injected into the \ac{CVS} for treatment of COVID-19 is presented. 
While the model accounts for propagation in branched topologies, only individual branchings are analyzed.
References~\cite{Chahibi2013,Mosayebi2019,Pal2023} highlight the lack of a 'network-wide' view of \ac{MC}. In particular, an abstract understanding of the impact that the network structure has on communication is missing in the literature. 
Along similar lines, the experimental \ac{MC} literature has repeatedly demonstrated the feasibility of communication via various signaling molecules in testbeds with simple channel structures such as single pipes or bifurcations~\cite{Bartunik2023,Wicke2022,Bartunik2022}. 
However, only very few testbeds with highly branched network topologies, reminiscent of those found in the \ac{CVS}, exist~\cite{Schaefer2024,Yu2024}.

To address these knowledge gaps, we here propose a generic analytical end-to-end model for the transport and reception of arbitrary signaling molecules in \acp{LBVN}.
Assuming particle transport can be characterized by linear systems and consequently \acp{CIR}, \acp{LBVN} approximate complex vessel networks like the \ac{CVS}~\cite{Chahibi2013}.
In particular, we extend the channel model in~\cite{Chahibi2013} to account for turbulent diffusion at the molecule injection site and network branching points, and link it to a transparent \ac{RX} model. 
Our primary focus is the investigation of the effects that the \ac{LBVN} topology exerts on the signal received at the outlet of the network.
We specialize the generic model to \acp{SPION} as information carriers~\cite{Palanisamy2019} and introduce a novel statistical model for reception via a planar coil \ac{RX}. 
These components, well established in previous testbeds~\cite{Bartunik2023,Bartunik2022}, harbor fundamentally fewer uncertainties compared to biological testbeds, e.g., based on fluorescent proteins or bacteria, allowing us to effectively isolate the impact of the topology.
Leveraging the generic model, we devise new metrics--namely the \textit{molecule delay} and the \textit{multi-path spread}--that relate the topology of any \ac{LBVN} to the degree of molecule dispersion induced by the network.
The degree of dispersion is in turn related to the \ac{SNR} of the signal received at the outlet of the network, enabling the characterization of the \ac{SNR} of any \ac{LBVN} \textit{solely based on its topology}.
In this manner, the proposed framework can be exploited to determine the optimal placement of sensors with certain \ac{SNR} requirements within highly branched vessel networks, such as the \ac{CVS}, or to predict the structural complexity allowed in testbeds to ensure sufficient signal strength at the~\ac{RX}.

The remainder of this paper is structured as follows:
Section~\ref{sec:SystemModel} provides the generic system model, which is specialized to a \ac{SPION}-based \ac{MC} system in Section~\ref{sec:SPIONTransportAndReception}. 
Building on this, Section~\ref{sec:BranchingMetrics} derives metrics that relate the topology of \acp{LBVN} to the \ac{SNR} at the \ac{RX}. 
The model is validated with testbed data and the impact of the topology of \acp{LBVN} is studied in Section~\ref{sec:Results}.
Final conclusions are presented in Section~\ref{sec:Conclusions}.

\section{System Model}\label{sec:SystemModel}
\begin{figure*}
	\centering
	\includegraphics[width=\textwidth]{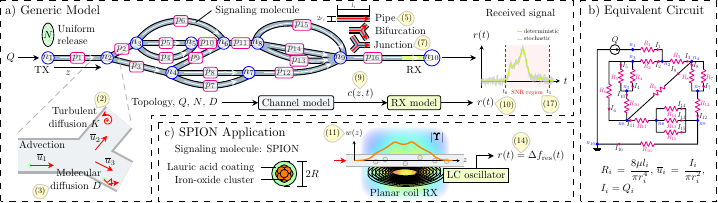}
	\caption{System model: a) $N$ signaling molecules are uniformly released in the cross-section of the network inlet (\ac{TX}), propagate through an \ac{LBVN} comprised of pipes, bifurcations, and junctions, influenced by advection as well as molecular and turbulent diffusion, and are received by a transparent \ac{RX}. b) The cross-sectional average flow velocity $\overline{u}_i$ in pipe $p_i$ is determined using an equivalent electrical circuit that models hydraulic resistance. c) \acp{SPION} as signaling molecules are received by a planar coil connected to an LC oscillator, yielding resonance frequency shift $\Delta f_\mathrm{res}(t)$ as received signal. The magnetic field around the coil is nonuniform and captured by the weighting function $w(z)$. Detailed explanations of the system components and notation are provided in the text, relevant equations are marked in yellow in the figure.}
	\label{fig:SystemModel}
\end{figure*}

In this section, we introduce a system model for the molecule release, i.e., the injection process, at the \ac{TX}, molecule propagation in the channel, and a transparent \ac{RX}, cf.~Fig.~\ref{fig:SystemModel}a). 
The system model is generically applicable to different types of signaling molecules and corresponding transparent \ac{RX} architectures.
In Section~\ref{sec:SPIONTransportAndReception}, the system model is specialized for the use of \acp{SPION} as information carriers.

\subsection{Molecule Injection}
At time $t=\SI{0}{\second}$, $N$ signaling molecules are injected instantaneously by the \ac{TX}, i.e., the inlet of the network at longitudinal coordinate $z=\SI{0}{\meter}$ of the corresponding vessel, cf.~Fig.~\ref{fig:SystemModel}a).
In practical systems, the injection via a Y-connector or venous cannula induces local turbulence in the fluid flow~\cite{Wicke2022}, which results in the immediate radial dispersion of the signaling molecules.
Accordingly, we postulate a uniform distribution of signaling molecules throughout the pipe cross-section from the moment of injection onwards.
The accuracy of this assumption is verified by the numerical results presented in Section~\ref{ssec:TestbedComparisonResults}.

\subsection{Linear Branched Vessel Networks}
Branched systems, such as the \ac{CVS}, generally exhibit complex topologies, comprising interconnected vessels of varying length, curvature, and irregular cross-sections. 
To facilitate analysis, these networks are often approximated using simplified models for their segments~\cite{Chahibi2013,Mosayebi2019,Pal2023}. 
Accordingly, we here consider \acp{LBVN} consisting of $E$ connected pipes, bifurcations, and $J$ junctions, as illustrated in~Fig.~\ref{fig:SystemModel}a).
\begin{enumerate}
\itemsep0em
	\item \textit{Pipe:} A pipe $p_i$ is a cylindrical vessel carrying fluid from its inlet to its outlet and is defined by its length $l_i$ and radius $r_i$. Pipes can be connected to other pipes, bifurcations or junctions at both the inlet and the outlet.
	\item \textit{Bifurcation:} A bifurcation is a connection point with no spatial extent, where an inflow pipe branches into several outflow pipes. Before and after each bifurcation, there must be connected pipes.
	\item \textit{Junction:} A junction $j_m$ is a connection point with no spatial extent, where several inflow pipes merge into one outflow pipe. Before and after each junction, there must be connected pipes. We denote the set of all inflow pipes of junction $j_m$ by $\mathcal{I}(j_m)$.
\end{enumerate}
Subsequently, we refer to the network inlet, outlet, and connection points, i.e., bifurcations and junctions, as nodes $n_v$, $\, v\in \left\lbrace 1,\ldots , V\right\rbrace $, where $V$ is the total number of nodes, and $n_1$ and $n_V$ are the nodes at the network inlet and outlet, respectively.
Pipes correspond to directed edges between these nodes, with the direction of the edge determined by the flow direction. 
This allows for a simplified representation of any \ac{LBVN} as a directed graph, as depicted with blue nodes and white arrows in Fig.~\ref{fig:SystemModel}a) for an exemplary network.
In addition, $\mathcal{P}(n_a,n_b)$ denotes the set of all paths between nodes $n_a$ and $n_b$ that differ in at least one pipe, with $a,b\in \left\lbrace 1,\ldots , V \right\rbrace$. 
Here, a path $P_k$ comprises a subset of connected pipes and junctions of the network and is denoted as
\begin{equation}\label{eqn:PathSet}
P_k=\left\lbrace p_i\,|\, i\in \mathcal{E}_k  \right\rbrace \cup \left\lbrace j_m\,|\, m\in \mathcal{J}_k \right\rbrace\,\text{,}
\end{equation}
where $\mathcal{E}_k\subseteq \left\lbrace 1,\ldots ,E \right\rbrace $ and $\mathcal{J}_k\subseteq \left\lbrace 1,\ldots ,J \right\rbrace $ are the index sets of the pipes and junctions in $P_k$, respectively.
Bifurcations are implicitly accounted for in $\mathcal{P}(n_a,n_b)$, as they determine which paths exist.

\subsection{Molecule Transport}\label{ssec:MoleculeTransport}
We restrict our model to advection-diffusion-based \ac{MC}. 
The flow rate $Q$ applied at the inlet gives rise to a background flow in all pipes of the network. 
While laminar flow prevails under typical conditions in the \ac{CVS} when considering \textit{individual pipes}~\cite{Wicke2022}, a fundamentally different flow behavior is associated with \acp{LBVN}. 
Here, turbulences at the injection site and connection points, like bifurcations and junctions, are expected to extend into the connected pipes, causing turbulent mixing throughout the majority of the highly branched network (cf. lower left part of Fig.~\ref{fig:SystemModel}a)). 
As a result, molecules are expected to be well-mixed in the cross-section at all times, permitting a (spatially) one-dimensional model for molecule transport~\cite{Chahibi2013}.
While modeling individual turbulences is intractable, a simple way to represent this type of molecular transport is by assuming underlying laminar flow in conjunction with eddy diffusion, represented by the eddy diffusion coefficient~\cite[Eq.~(4.16)]{Nieuwstadt2016}
\begin{equation}\label{eqn:TurbulentDiffusionCoefficient}
	K_i=\alpha \overline{u}_i r_i,\quad \alpha\in [0,2]\,\text{.}
\end{equation}
Here, $\alpha$ denotes a unitless proportionality constant.
In a given pipe $p_i$, $K_i$ is proportional to the distance orthogonal to the flow direction over which eddies persist, which is proportional to the pipe radius $r_i$ and flow velocity $\overline{u}_i$~\cite{Nieuwstadt2016}. 
The latter is obtained as follows:
For the typical parameters of the \ac{CVS}, see Section~\ref{ssec:InformationCarrier}, the model operates in the Aris-Taylor regime~\cite{Aris1956}.
This allows to simplify the modeling of the advection-diffusion process by averaging the radially dependent flow velocity inside any pipe $p_i$ to a one-dimensional cross-sectional average velocity $\overline{u}_i$, obtained from equivalent electrical circuits via node voltage analysis, see Fig.~\ref{fig:SystemModel}b) (for details see~\cite{Chahibi2013}). 
The corresponding flow rate in $p_i$ is given as $Q_i=\overline{u}_i\pi r_i^2$.

Besides this, molecules disperse due to molecular diffusion, characterized by the molecular diffusion coefficient~\cite[Eq.~(44)]{Chahibi2013}
\begin{equation}\label{eqn:MolecularDiffusionCoefficient}
	D=\dfrac{k_\mathrm{B}T}{6\pi\mu R}\,\text{,}
\end{equation}
where $k_\mathrm{B}=\SI{1.38e-23}{\meter\squared\kilo\gram\per\second\squared\per\kelvin}$, $T$, $\mu$, and $R$ denote the Boltzmann constant, the fluid temperature, the dynamic fluid viscosity, and the molecule radius, respectively.
Molecular and eddy diffusion can be modeled as additive phenomena and are both incorporated in the total diffusion coefficient $D^\mathrm{tot}_i=D+K_i$.
Lastly, the Aris-Taylor effective diffusion coefficient is given as~\cite[Eq.~(26)]{Aris1956}
\begin{equation}\label{eqn:ArisTaylorEffectiveDiffusionCoefficient}
	D^\mathrm{eff}_i=\dfrac{r_i^2 \overline{u}_i^2}{48 D^\mathrm{tot}_i}+D^\mathrm{tot}_i\,\text{.}
\end{equation}

On the basis of the above models, molecule transport in \acp{LBVN} can be described using \acp{CIR}~\cite{Chahibi2013}.
Specifically, a \ac{CIR} can be derived for each of the three types of channel segments. 
Under the assumption of linearity, the end-to-end network \ac{CIR} then results from the convolution and superposition of corresponding segment \acp{CIR} and molecule fluxes.

Solving the advection-diffusion equation results in the \ac{CIR} of pipe $p_i$ in unit $\SI{}{\per\meter}$~\cite[Eq.~(49)]{Chahibi2013}
\begin{equation}\label{eqn:SinglePipeCIR}
	h_i(z,t)\hspace*{-.5mm}=\hspace*{-.5mm}\dfrac{1}{\sqrt{4\pi D^\mathrm{eff}_it}}\exp\hspace*{-.5mm} \left(\hspace*{-1mm}-\dfrac{(z-\overline{u}_it)^2}{4D^\mathrm{eff}_it}\hspace*{-.5mm} \right)\hspace*{-1mm},\,t\hspace*{-1mm}>\hspace*{-1mm}0,z\hspace*{-1mm}\in\hspace*{-1mm} [0,l_i]\,\text{,}
\end{equation}
where $z$ denotes the longitudinal coordinate \textit{within the pipe}, i.e., reaching from its inlet at $z=0$ to its outlet at $z=l_i$. 
The instantaneous net molecule flux for $N=1$ at position $z$ in pipe $p_i$ comprises a diffusive and an advective term, is given in unit $\SI{}{\per\second}$, and follows from~(\ref{eqn:SinglePipeCIR}) as~\cite[Eq.~(4.4)]{Berg1993}
\begin{equation}\label{eqn:PipeFlux}
		J_i(z,t)\hspace*{-.8mm}=\hspace*{-.8mm} -D^\mathrm{eff}_i\dfrac{\partial h_i(z,t)}{\partial z} \hspace*{-.8mm}+\hspace*{-.8mm} \overline{u}_i h_i(z,t)
		 \hspace*{-.8mm}=\hspace*{-1mm}\left( \hspace*{-1mm}\dfrac{z - \overline{u}_i t}{2t} \hspace*{-.8mm}+ \hspace*{-.8mm}\overline{u}_i\hspace*{-1mm}\right)\hspace*{-1mm}h_i(z,t)\,\text{.}
\end{equation}
The principle of mass conservation yields the \ac{CIR} of a junction $j_m$ that is traversed through $p_i$~\cite[Eq.~(22)]{Mosayebi2019}
\begin{equation}\label{eqn:JunctionCIR}
	h^i_m(t)=\dfrac{Q_{i}}{\sum_{k\in\mathcal{I}(j_m)}Q_k}\delta (t)\,\text{,}
\end{equation}
i.e., molecules partition according to the ratio of flow rates. 
Here, $\delta (\cdot)$ denotes the Dirac delta function.
The end-to-end \ac{CIR} of the network between $n_a$ and $n_b$ equals the sum of the individual path \acp{CIR}, which in turn result from the convolution of the molecule flux entering the \ac{RX} pipe $p_{i'}$ and its \ac{CIR}
\begin{multline}\label{eqn:EndToEndCIR}
		h_{n_a,n_b}(z,t)=\\ \hspace*{-2.2mm}\left[ \sum\limits_{P_k\in\mathcal{P}(n_a,n_b)} \left(\Conv_{ \substack{i\in \mathcal{E}_k:\\ p_i\neq p_{i'}}} \hspace*{-3mm} J_i(l_i,t)\hspace*{2mm} * \hspace*{-2.5mm}\Conv_{\substack{(i,m)\in \mathcal{E}_k\times \mathcal{J}_k:\\ p_i\in \mathcal{I}(j_m) }} \hspace*{-9mm} h^ i_m(t)  \right)  \right] * h_{i'} (z,t)\,\text{,}
\end{multline}
where $*$ and $\Conv_{i\in\mathcal{S}\subset\mathbb{N}}J_i=J_{i_1}*J_{i_2}*\ldots *J_{i_{\vert \mathcal{S}\vert}}$ denote the convolution operator and the convolution operator over a set with respect to time, respectively. Here, $\mathbb{N}$ and $\vert \cdot \vert$ denote the set of natural numbers and the cardinality of a set, respectively.
Since molecule concentration remains unchanged across bifurcations~\cite[Eq.~(53)]{Chahibi2013}, their \acp{CIR} do not appear in~(\ref{eqn:EndToEndCIR}).
From~(\ref{eqn:EndToEndCIR}), the received molecule concentration in unit $\SI{}{\per\meter}$ is obtained as
\begin{equation}
	c(z,t)= N h_{n_a,n_b}(z,t)\,\text{.}
\end{equation}

\subsection{Molecule Reception}
We consider a transparent \ac{RX} whose domain extends over $z$. The generic received signal results from the concentration~as 
\begin{equation}\label{eqn:ReceivedSignal}
	r(t)=f\left( \int_{z\in\mathrm{dom}(w(\cdot))} w(z)c(z,t)\,\mathrm{d}z\right) \,\text{,}
\end{equation}
where $\mathrm{dom}(\cdot)$ denotes the domain operator. 
Here, the weighting function $w(z)$ and the signal conversion function $f(x)$ are specific to the sensor employed as \ac{RX}. 
Thereby, potential inhomogeneities of the sensing process\footnote{Consider, e.g., optical \acp{RX} in the \ac{CVS} where varying refractive indices of surrounding tissue cause sensing inhomogeneities over space.} over $z$ are captured by $w(z)$ and subsequent processing steps\footnote{Consider, e.g., differential \acp{RX} where the current measured signal amplitude is compared to that measured in previous time steps or a reference signal.} are mapped by $f(x)$.

In the \ac{MC} literature, often the special case of a so-called perfect counting \ac{RX}, extending from $z^\mathrm{-}$ to $z^\mathrm{+}$ with $w(z)=1$, $\forall z\in [z^\mathrm{-},z^\mathrm{+}]$, and linear mapping, i.e., $f(x)=x$, is assumed. 
In this case, the received signal equals the number of observed molecules $N^\mathrm{obs}(t)=\int_{z^\mathrm{-}}^{z^\mathrm{+}}c(z,t)\,\mathrm{d}z$.
In practice, however, sensors with such properties do not exist, emphasizing the need for the sensor-related degrees of freedom in~(\ref{eqn:ReceivedSignal}).

\section{SPION Transport and Reception}\label{sec:SPIONTransportAndReception}
In this section, we describe how the system model introduced above can be adapted to \ac{SPION}-based \ac{MC}. To this end, we first present the physical parameters of \acp{SPION} and derive the resulting transport properties. Secondly, a novel analytical model for the reception process using a planar coil is proposed.

\subsection{Information Carriers}\label{ssec:InformationCarrier}
\acp{SPION} are superparamagnetic nanoparticles, e.g., synthesized through co-precipitation~\cite{Palanisamy2019}. 
At their core, they hold a cluster of iron-oxide molecules, responsible for a high magnetic susceptibility $\chi_\mathrm{ref}\approx \SI{3e-3}{}$ (SI units)~\cite{Wicke2022}. 
As a result, \acp{SPION} become strongly magnetized in the presence of an external magnetic field, but show no remanence once removed from the field, due to their small size~\cite{Bartunik2023}.
Various coatings can be applied to the core, including biocompatible coatings like human serum albumin, chemical degradation and agglomeration mitigating coatings like Dextran, therapeutic coatings carrying drug particles, or ligands that bind to desired target sites~\cite{Palanisamy2019}.
Depending on the coating, the diameters of \acp{SPION} range from below $\SI{10}{\nano\meter}$ to micrometers~\cite{Palanisamy2019}.
In the following, we consider \acp{SPION} with a lauric acid coating for chemical stability and biocompatibility having a radius of $R=\SI{24.5}{\nano\meter}$~\cite{Wicke2022}.

\subsection{Diffusion}
For an assessment of the dynamic properties of \acp{SPION}, consider the following scenario. 
To enable comparison with testbed data, hereafter we assume that the molecules are suspended in distilled water at room temperature $T=\SI{293}{\kelvin}$ with $\mu = \SI{1e-3}{\kilo\gram\per\meter\per\second}$.
This results in a molecular diffusion coefficient $D\approx\SI{8.61e-12}{\meter\squared\per\second}$, cf.~(\ref{eqn:MolecularDiffusionCoefficient}). 
Typical medium-sized arteries in the human \ac{CVS}, e.g., in the index finger, exhibit a radius in the order of $r\approx\SI{6.5e-4}{\meter}$ and an average flow velocity around $\overline{u}\approx \SI{5e-2}{\meter\per\second}$~\cite{Klarhoefer2001}. 
According to~(\ref{eqn:TurbulentDiffusionCoefficient}), this yields a maximum eddy diffusion coefficient of $K\approx \SI{6.5e-5}{\meter\squared\per\second}\gg D$ for $\alpha=2$. 
As the exact value of $\alpha$ can typically only be obtained experimentally, in Section~\ref{ssec:TestbedComparisonResults}, we fit $\alpha$ to testbed data from~\cite{Bartunik2023}. 
Lastly, utilizing~(\ref{eqn:ArisTaylorEffectiveDiffusionCoefficient}) yields a maximum effective diffusion coefficient of $D^\mathrm{eff}\approx \SI{6.53e-5}{\meter\squared\per\second}$ ($\alpha=2$).

\subsection{Planar Coil Receiver}
Due to their magnetic properties, \acp{SPION} can be detected using capacitive and inductive \acp{RX}. 
In this work, we model the inductive \ac{RX} used in the testbed in~\cite{Bartunik2023}. 
The testbed comprises a straight channel pipe, molecule injection via a venous cannula, and a planar coil \ac{RX}\footnote{The testbed employs coil "K" from the LDCCOILEVM reference coils board by Texas Instruments, cf.~\url{https://www.ti.com/tool/LDCCOILEVM}.} with a connected LC oscillation circuit.
Unlike conventional cylindrical coils, planar coils can be positioned adjacent to the vessels, eliminating the need to envelop them, cf.~Fig.~\ref{fig:SystemModel}c). 
When \acp{SPION} enter the vicinity of the coil, a cascade of physical processes leads to the detection of the molecules through the oscillator, as detailed in the following.

\textit{1) Magnetic Susceptibility:} In the presence of \acp{SPION}, the volume magnetic susceptibility $\chi_\mathrm{v}$ (unitless) of the environment increases, since water and air exhibit much lower susceptibilities compared to the nanoparticles.
Given that the magnetic field around planar coils is non-homogeneous, cf.~Fig.~\ref{fig:SystemModel}c), different \acp{SPION} contribute differently to the received signal, depending on their position.
In particular, their contribution is proportional to the strength of the magnetic field\footnote{For simplicity, we assume the \ac{RX} pipe diameter is small enough, such that all \acp{SPION} at a given $z$ experience the same magnetic field strength.}, denoted by $|\pmb{\Upsilon} (z)|$, which we capture using the unitless weighting function~\cite{Wicke2022}
\begin{equation}\label{eqn:CoilWeightingFunction}
	w(z)=\beta \chi_\mathrm{ref} |\pmb{\Upsilon}(z)|\,\text{,}
\end{equation}
where $\beta$ denotes a proportionality constant. 
Since closed-form expressions for the magnetic field strength can only be found for simple coil geometries, we obtain $|\pmb{\Upsilon}|$ through COMSOL\textsuperscript{\textregistered} simulations\footnote{\label{foot:COMSOLSimulationFile}For the simulation file, see~\url{https://doi.org/10.5281/zenodo.13744883}.}. 
Subsequently, the volume magnetic susceptibility $\chi_\mathrm{v}(t)$ is obtained from the integral in~(\ref{eqn:ReceivedSignal}), i.e., $\chi_\mathrm{v}(t)=\int_{z\in\mathrm{dom}(w(\cdot ))}w(z)c(z,t)\,\mathrm{d}z$, using the weighting function in~(\ref{eqn:CoilWeightingFunction}).

\textit{2) Inductance:} A change in $\chi_\mathrm{v}(t)$ in turn implies a change in the coil inductance
\begin{equation}\label{eqn:MSPCInductance}
	L(t)=\mu_\mathrm{r}(t)L_0=\left(1+\chi_\mathrm{v}(t) \right)L_0\,\text{,}
\end{equation}
where $\mu_\mathrm{r}(t)$ is a unitless quantity denoting the relative permeability and $L_0$ denotes the coil inductance when no \acp{SPION} are in the proximity. 
COMSOL simulations\footnoteref{foot:COMSOLSimulationFile} yield $L_0=\SI{206.51}{\micro\henry}$, which is very close to the value $L_0=\SI{206.227}{\micro\henry}$ reported by the coil manufacturer, which we use in the following. 

\textit{3) Resonance Frequency Shift:} The LC oscillation circuit in~\cite{Bartunik2023} employs a capacitor with capacitance $C=\SI{68}{\pico\farad}$ and resonates at frequency
\begin{equation}\label{eqn:NaturalResonanceFrequency}
	f_\mathrm{res}(t)=\dfrac{1}{2\pi \sqrt{L(t)C}}\,\text{.}
\end{equation}
Let $f_{\mathrm{res},0}$ denote the resonance frequency if no \acp{SPION} are near the coil, i.e., for $L(t)=L_0$. 
Then, we define the deterministic received signal as resonance frequency shift~\cite[Eq.~(1)]{Bartunik2022}
\begin{equation}\label{eqn:ResonanceFrequencyShift}
	\Delta f_\mathrm{res}(t)=f_\mathrm{res}(t)-f_{\mathrm{res},0}
=\dfrac{\sqrt{L(t)}-\sqrt{L_0}}{2\pi\sqrt{L_0L(t)C}}\,\text{.}
\end{equation}
Consequently, complying with the generic framework, the received signal $r(t)$ for \ac{SPION}-based \ac{MC} results from~(\ref{eqn:ReceivedSignal}) with $w(z)$ in~(\ref{eqn:CoilWeightingFunction}) and $f(x)$ comprised of~(\ref{eqn:MSPCInductance}) -- (\ref{eqn:ResonanceFrequencyShift}).

In practice, the received signal is impacted by noise. 
To characterize the noise at the planar coil \ac{RX}, we evaluate the testbed \acp{CIR} in~\cite{Bartunik2023}, measured at varying channel lengths and background flow rates in terms of resonance frequency shift in unit $\SI{}{\hertz}$. 
For each testbed parameter setting, we collect the deviations between ten measured \acp{CIR} and their respective ensemble-averaged \ac{CIR} over time, and analyze the distribution of the deviations. 
By focusing on a low-signal amplitude time window around the \ac{CIR} tails, we observe similar signal-independent, additive white Gaussian noise $n_\mathrm{sensor}$ across all settings, which is accurately modeled~as
\begin{equation}
	n_\mathrm{sensor}\sim \mathcal{N}(\xi =\SI{0}{\hertz},\sigma^2=\SI{0.3822}{\kilo\hertz\squared})\,\text{,}
\end{equation}
where $\mathcal{N} (\xi,\sigma^2)$ denotes the Gaussian probability density function with mean $\xi$ and variance $\sigma^2$.
This noise, attributed to sensor inaccuracies, is the most prominent noise affecting the data, leading to the statistical received signal model
\begin{equation}\label{eqn:StatisticalReceivedSignal}
		\Delta F_\mathrm{res}(t)=\Delta f_\mathrm{res} (t) + n_\mathrm{sensor}(t)\,\text{,}
\end{equation}
i.e., for any $t$, $\Delta F_\mathrm{res}(t)$ is a random variable. 
Similar to~\cite[Eq.~(5)]{Jamali2017}, we define a time-average \ac{SNR}. 
In particular, for the averaging, we capture the entire received signal, excluding its vanishing onset and tail, cf.~right part of Fig.~\ref{fig:SystemModel}a),
\begin{equation}\label{eqn:SNR}
	\begin{aligned}
	&\mathrm{SNR}=10\log_{10}\left( \dfrac{\frac{1}{t_1-t_0}\int_{t_0}^{t_1} \Delta f_\mathrm{res}^2(t)\,\mathrm{d}t}{\sigma^2}\right),\\
	& t_0=\sup \{ t < t_\mathrm{first\, path} \,|\, \Delta f_\mathrm{res}(t) \leq \epsilon \},\\
		& t_1=\inf \{ t > t_\mathrm{last\, path} \,|\, \Delta f_\mathrm{res}(t)  \leq \epsilon \}\,\text{,}
	\end{aligned}
\end{equation}
where $t_\mathrm{first\, path}$ and $t_\mathrm{last\, path}$ denote the first and last path peak times (cf.~(\ref{eqn:PathTmax})) in $\Delta f_\mathrm{res}(t)$, respectively, and $\epsilon$ denotes a small threshold, chosen as $\epsilon =\SI{1}{\milli\hertz}$ in our simulations.

\section{Topology-Dispersion Metrics}\label{sec:BranchingMetrics}
To characterize how the network topology affects the received molecular signal, below, we propose metrics that relate the topology of \acp{LBVN} to the received \ac{SNR}. 
In particular, these metrics are derived from the generic system model in Section~\ref{sec:SystemModel} and characterize how strongly molecules are dispersed by a given network topology, taking into account molecular and eddy diffusion. 
The degree of dispersion is in turn associated with the received \ac{SNR}. 
This methodology is akin to the characterization of multi-path channels in wireless communications, where similar metrics are employed to assess the influence of the channel on the quality of the received signal~\cite{Rappaport2024}.

\textit{1) Molecule Delay:} 
The extent of molecule dispersion via diffusion from the network inlet to the outlet is inherently linked to the molecule travel time, with longer delays implying larger dispersion, cf.~(\ref{eqn:SinglePipeCIR}).
Our first metric below captures this.

For a single pipe $p_i$, the time when most molecules are located at its outlet, denoted as the pipe peak time, is given as
\begin{equation}\label{eqn:PipeTmax}
	t^\mathrm{peak}_{p_i}=\argmax_t h_i(l_i,t)=\dfrac{-D^\mathrm{eff}_i+\sqrt{{D^\mathrm{eff}_i}^2+\overline{u}_i^2l_i^2}}{\overline{u}_i^2}\,\text{.}
\end{equation}
From~(\ref{eqn:PipeTmax}), the time when most molecules are located at the outlet of a path $P_k$, denoted as the path peak time, follows~as
\begin{equation}\label{eqn:PathTmax}
	t^\mathrm{peak}_{P_k}=\sum\limits_{p_i\in P_k} t^\mathrm{peak}_{p_i}\,\text{.}
\end{equation}
Moreover, the fraction of molecules traveling through $P_k$ is
\begin{equation}\label{eqn:PathGamma}
	\gamma_{P_k} = \prod\limits_{\substack{p_i,j_m\in P_k\\ p_i\in\mathcal{I}(j_m )}} \dfrac{Q_{i}}{\sum_{p_v\in\mathcal{I}(j_m)} Q_{v}}\,\text{,}
\end{equation}
i.e., the product of the flow rate fractions of all junctions contained in the path. 
From~(\ref{eqn:PipeTmax}) -- (\ref{eqn:PathGamma}), we define the \textit{molecule delay} for a given network between nodes $n_a$ and $n_b$ as
\begin{equation}\label{eqn:NetworkAverageTimeOfArrivalWeightedAverage}
	t_{n_a}^{n_b}=\sum\limits_{P_k\in \mathcal{P}(n_a,n_b)} \gamma_{P_k}t^\mathrm{peak}_{P_k}\,\text{,}
\end{equation}
giving, via $\gamma_{P_k}$, more weight to paths carrying more molecules.
Note that $t_{n_a}^{n_b}$ resembles the \textit{excess delay} in multi-path wireless communications, quantifying the mean signal delay~\cite{Rappaport2024}.

\textit{2) Multi-Path Spread:}
Secondly, networks with paths having widely varying peak times $t^\mathrm{peak}_{P_k}$ lead to increased dispersion, causing molecules to arrive at the \ac{RX} more spread out over time.
We capture this phenomenon by the \textit{multi-path spread}
\begin{equation}\label{eqn:PTV}
	\sigma_{n_a}^{n_b}=\sqrt{\sum_{P_k\in\mathcal{P}(n_a,n_b)} \gamma_{P_k} \left(t^\mathrm{peak}_{P_k}-t_{n_a}^{n_b} \right)^2}\,\text{.}
\end{equation}
Note that $\sigma_{n_a}^{n_b}$ parallels the \textit{root mean square delay spread}, describing signal spread in a multi-path wireless channel~\cite{Rappaport2024}.
\begin{figure}
	\centering
	\includegraphics[scale=1]{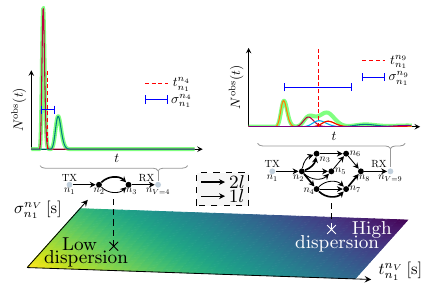}
	\caption{The position of any \ac{LBVN} in the dispersion space is solely based on its topology, as $t_{n_1}^{n_V}$ and $\sigma_{n_1}^{n_V}$ characterize the dispersion of the molecules propagating from \ac{TX} to \ac{RX}. Two exemplary \acp{LBVN} with identical pipe radii and pipe lengths of either $l$ or $2l$, along with their received signals $N^\mathrm{obs}(t)$, are shown. Thin colored curves show individual path contributions; the green curve shows the total signal.}
	\label{fig:Metrics}
\end{figure}

\textit{3) Dispersion Space:}
$t_{n_a}^{n_b}$ and $\sigma_{n_a}^{n_b}$ span a two-dimensional space we term \textit{dispersion space}, cf.~Fig.~\ref{fig:Metrics}.
Any \ac{LBVN} can be located in the space \textit{solely based on its topology}.
Setting $n_a=n_1$, $n_b=n_V$, the position in the space can then be used to infer the degree to which molecules disperse while propagating from \ac{TX} to \ac{RX}.
\acp{LBVN} close to the origin experience comparatively little dispersion and \acp{LBVN} far away from the origin suffer from strong dispersion, cf.~left and right network in Fig.~\ref{fig:Metrics}, respectively. 
Lastly, we hypothesize that the level of dispersion correlates negatively with the received \ac{SNR}.

\section{Numerical Evaluation}\label{sec:Results}
In this section, we validate our end-to-end model with testbed data and subsequently evaluate how the dispersion space can be used to infer the received \ac{SNR} from the network topology.

\subsection{Comparison to Testbed Data}\label{ssec:TestbedComparisonResults}
\begin{figure}
	\centering
	\includegraphics[width=.5\textwidth]{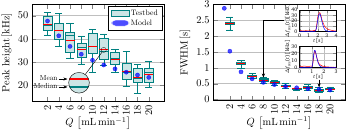}
	\caption{Comparison of testbed \acp{CIR} in~\cite{Bartunik2023} and model \acp{CIR} at varying flow rates $Q$ in terms of peak height and \ac{FWHM}.}
	\label{fig:ValidationWithTestbedData}
\end{figure}

For the validation of our end-to-end model, we compare the \acp{CIR} measured in~\cite{Bartunik2023} with those predicted by~(\ref{eqn:ResonanceFrequencyShift}), focusing on \ac{CIR} peak height and the \ac{CIR} \ac{FWHM}, cf.~Fig.~\ref{fig:ValidationWithTestbedData}. 
These quantities are crucial for signal reception~\cite{Bartunik2023}.
We adopt all default parameters from~\cite[Tab.~1]{Bartunik2023}, vary the background flow rate $Q$ for a fixed channel length $l=\SI{5}{\centi\meter}$ as done in~\cite{Bartunik2023}, and estimate $\alpha =0.01$ and $\beta = \SI{5e-12}{}$ by fitting our model to the testbed data.
In~\cite{Bartunik2023}, ten \ac{CIR} realizations are measured for each setting, the statistics of which are illustrated in Fig.~\ref{fig:ValidationWithTestbedData} using boxplots.

First, we consider the peak heights of the \acp{CIR}. 
Testbed and model data show good agreement, with decreasing peak heights as $Q$ increases.
This trend occurs because greater $Q$ imply greater flow velocities and thus stronger effective diffusion, cf.~(\ref{eqn:ArisTaylorEffectiveDiffusionCoefficient}), which ultimately leads to molecules being more spread out over space, i.e., less \acp{SPION} are present in the magnetic field of the \ac{RX} at pipe peak time.
The deviations between $\SI{10}{}$ and $\SI{14}{\milli\liter\per\minute}$ are likely due to testbed measurement inaccuracies, as these data points significantly deviate from the general trend.
Secondly, the model \ac{FWHM} aligns with the testbed data, decreasing as $Q$ increases because molecules pass through the magnetic field of the \ac{RX} more quickly at higher $Q$, resulting in narrower \acp{CIR}.
Notable deviations for low $Q$ are likely due to channel soiling, which becomes more pronounced in the testbed at low flow velocities and is not accounted for in our model.
In summary, these results demonstrate that our model accurately reproduces the testbed data.

\subsection{Dispersion Space}
\begin{figure}
	\centering
	\includegraphics[width=.48\textwidth]{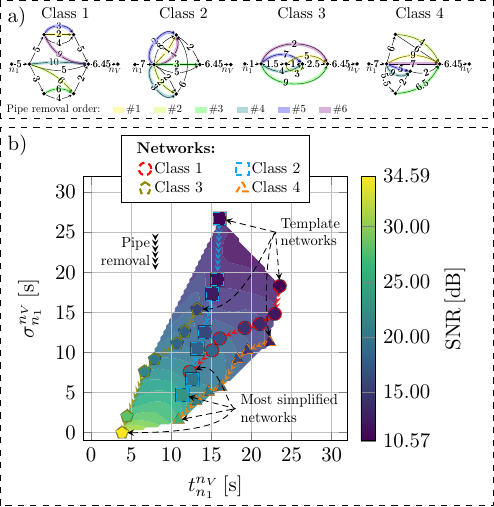}\vspace*{1.5mm}
	\caption{Dispersion space filled with \acp{LBVN}. a) 28 networks are generated from four template \acp{LBVN} by removing one pipe per iteration in the indicated removal order. Pipe lengths in $\SI{}{\centi\meter}$ are given as edge weights, with $r=\SI{1}{\milli\meter}$. b) Markers correspond to different \acp{LBVN} and color-code the \ac{SNR} at the network outlets. The clear pattern between location in the space and \ac{SNR} enables the characterization of the \ac{SNR} based on the topology alone.}
	\label{fig:DispersionSpace}
\end{figure}

To verify that the received \ac{SNR} can be directly inferred from the network topology utilizing the proposed dispersion space, we analyze a variety of synthetically generated \acp{LBVN}. 
For each of the four template networks shown in Fig.~\ref{fig:DispersionSpace}a), six further networks are generated by iteratively removing one pipe in the order indicated in color, totaling 28 networks. 
The pipe lengths in unit $\SI{}{\centi\meter}$ are given as edge weights in Fig.~\ref{fig:DispersionSpace}a), with a radius of $r=\SI{1}{\milli\meter}$.
In each network, we simulate the instantaneous injection of $N=\SI{2e12}{}$ \acp{SPION} (as done in~\cite{Bartunik2023}) at a flow rate of $Q=\SI{1e-7}{\meter\cubed\per\second}$ and calculate the received \acp{SNR} as described in~(\ref{eqn:SNR}).
Additionally, we map the networks within the dispersion space based on their topologies, with the \acp{SNR} color-coded and the removal of pipes indicated by arrow paths, cf.~Fig.~\ref{fig:DispersionSpace}b). 
For the sake of visualization, we interpolate the \ac{SNR} in the spaces between the networks.

Fig.~\ref{fig:DispersionSpace}b) reveals a strong correlation and clear pattern between the position of each \ac{LBVN} in the space and its corresponding \ac{SNR}, confirming our expectations from Section~\ref{sec:BranchingMetrics}.
Indeed, we find that networks exhibit systematically weaker \acp{SNR} with increasing $t_{n_1}^{n_V}$ and $\sigma_{n_1}^{n_V}$ due to enhanced dispersion, validating our choice of metrics. 
For example, the most simplified network from class 3 is the only \ac{LBVN} with a single path from $n_1$ to $n_V$, resulting in $\sigma_{n_1}^{n_V}=0$ and the highest \ac{SNR}.
Class 4 networks predominantly feature similar paths, leading to low $\sigma_{n_1}^{n_V}$, while the class 2 template exhibits paths of significantly different lengths, resulting in its simplifications being distributed across a wide range of $\sigma_{n_1}^{n_V}$.
Despite these distinct class characteristics, our metrics capture the underlying toplogy-\ac{SNR} relation.

\section{Conclusion}\label{sec:Conclusions}
In this paper, we investigated how the topology of an \ac{LBVN} affects the signal received at the outlet of the network. 
To allow quantitative statements, we introduced a generic end-to-end model for diffusion-advection-driven \ac{MC} in \acp{LBVN}.
The model extends an existing channel model to include turbulences induced at the injection site and at the branching points, and links it to a transparent \ac{RX} model. 
We adapted the generic model for communication using \acp{SPION} to effectively isolate the effects topology imposes and to validate the end-to-end model with existing testbed data.
Along with this, we presented a novel analytical model for the reception of \acp{SPION} using a planar coil.
Furthermore, we proposed two metrics, namely \textit{molecule delay} and \textit{multi-path spread}, that relate the network topology to molecule dispersion and thus to the \ac{SNR} at the network outlet.
This allows to characterize the received \ac{SNR} solely based on the topology.
Consequently, our framework can, e.g., aid optimal sensor placement in systems like the \ac{CVS} under \ac{SNR} constraints or predict branched testbed topologies that exhibit both structural complexity and reliable signal detection at the sensor.

Moving forward, we intend to further validate the \ac{SPION}-based model with measurements from highly branched testbeds.
Similarly, the modeling and comparison of other signaling molecule-\ac{RX} pairs, e.g., based on ink particles or fluorescent proteins and spectral sensors, may be insightful.

\renewcommand{\baselinestretch}{1}
\bibliography{references}

\end{document}